\input{psfig.sty} 

\documentstyle[editedvolume,psfig]{crckapb}

\def\spose#1{\hbox to 0pt{#1\hss}}
\def\simlt{\mathrel{\spose{\lower 3pt\hbox{$\mathchar"218$}}
     \raise 2.0pt\hbox{$\mathchar"13C$}}}
\def\simgt{\mathrel{\spose{\lower 3pt\hbox{$\mathchar"218$}}
     \raise 2.0pt\hbox{$\mathchar"13E$}}}

\def\ie{{\rm i.e. }}
\def\etal{{\rm et~al. }}

\begin{opening}
\title{Stellar Yields and Chemical Evolution}

\author{Brad K. Gibson}
\institute{Mount Stromlo \& Siding Spring Observatories\\
           Weston Creek P.O., Weston, ACT, 2611  Australia}

\end{opening}

\runningtitle{Stellar Yields \& Chemical Evolution}

\begin{document}


\section{Introduction}

Several speakers at this symposium have alluded to the zeroth-order agreement
between the Type II supernovae (SNe) stellar yields, as predicted by the models
of those most responsible for driving progress in 
the field - \ie Arnett (1991,1996);
Maeder (1992); Woosley \& Weaver (1995); Langer \& Henkel (1995); Thielemann 
\etal (1996), hereafter referred to as A91, A96, M92, WW95, LH95, and TNH96,
respectively.  It is important though for those entering (or indeed, already
involved in!) the chemical evolution field to be cognizant of the fact that
there are important first- and second-order differences between the yield
compilations.

In the next few pages, I provide
a qualitative comparison of the currently available Type II SNe yield grids.
The strengths and weaknesses of a given grid, demonstrated by comparing against
relevant observations, are noted.  Some simple chemical evolution models are
shown which graphically demonstrate the effect of yield grid selection.

\section{Stellar Yields}

The most recent oxygen and iron yield predictions, from the aforementioned
modelers, are shown in Figure 1.  While the general trend of oxygen ejected as
a function of progenitor mass is similar -- the solar metallicity M92 yields
are the exception, due primarily to the inclusion of metallicity-dependent 
mass-loss
in the evolutionary models -- it is also apparent that there
is a zero-point uncertainty at the level of a factor of $\sim 2\rightarrow 3$.
Because of remaining uncertainties in the treatment of convection, mass-loss,
reaction rates, evolving helium cores versus self-consistent evolution from the
zero age main sequence (ZAMS), etc, we simply cannot (yet!)
predict oxygen yields to an accuracy better than this factor of $\sim
2\rightarrow 3$ (Langer 1997).  

While the predicted oxygen yield depends primarily upon the physics of
hydrostatic burning, that of iron is more closely tied to that of explosive
nucleosynthesis.  Iron is more problematic than oxygen in many ways as it is
linked inexorably to the placement of the mass-cut, effectively a free
parameter in the stellar models.  This makes the iron yields of Figure 1 \it
highly \rm uncertain; other than the ``calibration'' points provided by SNe
1987A and 1993J (see TNH96) at $m=20$ M$_\odot$ and 14 M$_\odot$, respectively,
\it a priori \rm predictions of iron yields at other progenitor masses should
be regarded with caution.

\begin{figure}[htbp]
\centerline{
\psfig{file=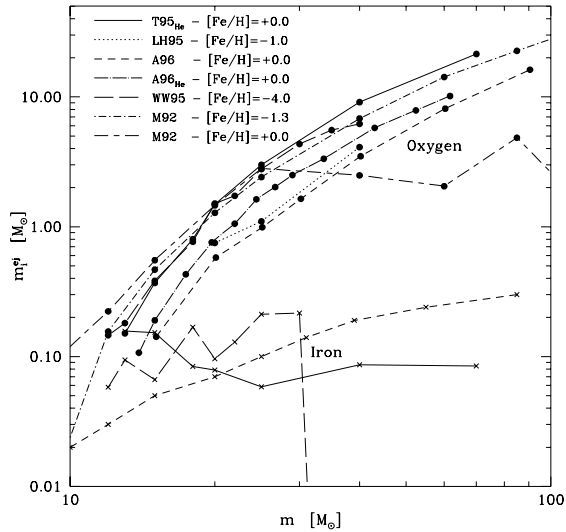,width=3.00in,height=3.00in,angle=0}}
\caption{Oxygen and iron yields as a function of progenitor mass, for the
primary Type II SNe yield compilations currently available.
The models of M92, LH95, and A96 are only
evolved to the completion of oxygen burning; those with a subscript `He'
appended refer to evolved helium cores only - the remainder were evolved
self-consistently from the ZAMS.}
\end{figure}

Figure 2 shows the ejecta C/O ratio, as a function of progenitor mass, for 
a variety of model sources.  Whereas oxygen by itself has a factor of $\sim 3$
uncertainty, the ratio of C/O is even worse.  Here we can see that for $m\simgt
15$ M$_\odot$, the uncertainty has grown to a factor of $\sim 5\rightarrow
10$!  The shaded region represents the Reimers' et al (1992) claim of log
C/O$\approx -1$ in high redshift Lyman-limit systems.  It should be readily
apparent that, regardless of initial mass function (IMF) and star formation
history, the A91 and LH95 yields would be hard-pressed to ever recover such a
low C/O.

\begin{figure}[htbp]
\centerline{
\psfig{file=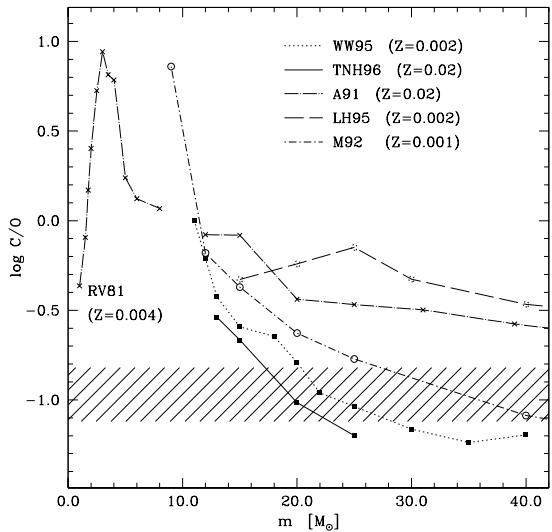,width=3.00in,height=3.00in,angle=0}}
\caption{Ratio of carbon to oxygen in the model Type II SNe yields.  Renzini \&
Voli's (1981) low and intermediate mass stellar yields are shown for
comparison.  The shaded region represents the observed C/O in high-redshift
Lyman-limit systems (Reimers \etal 1992).}
\end{figure}

Figure 3 shows the predicted yield [O/Mg] and [Si/Mg], as a function of
progenitor mass, overlaid on the observed Milky Way bulge giant underabundances
(\ie [O,Si/Mg]$\approx -0.3$), as reported by Rich (1996).  Again, independent
of IMF or star formation history, we can conclude that the WW95 solar
metallicity yields are incompatible with Rich's (1996) bulge [O/Mg].  The TNH96
[O/Mg] are only marginally compatible, as are the WW95 [Si/Mg].  Resorting to
Type Ia SNe to alleviate the ``discrepancy'' does not help as [O,Si/Mg]$_{\rm
Ia}>+0.0$.

\begin{figure}[htbp]
\centerline{
\psfig{file=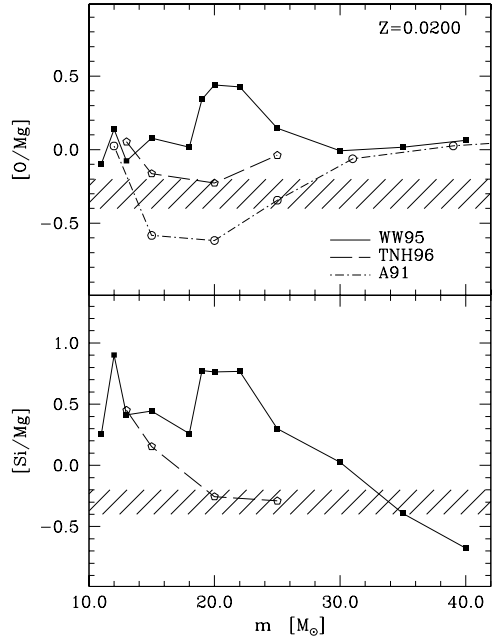,width=3.45in,height=3.45in,angle=0}}
\caption{Comparison of the model Type II SNe yields, with the Milky Way bulge
giant abundances (shaded region), as noted in Rich (1996).}
\end{figure}

\section{Chemical Evolution}

Figure 4 demonstrates how the different yield behaviors alluded to previously
translate into a simple chemical evolution application.  Here I show the
evolution of the interstellar medium metallicity Z$_{\rm g}$, C/O, and [O/Fe]
for a massive elliptical galaxy, assuming star formation is directly
proportional to the available gas mass.  The star formation timescale has been
adjusted to ensure the present-day photo-chemical properties of the models
match local observations.  All other input ingredients are identical.  Despite
the final (V-K) and $<$Z$>_\ast$ being the same in all cases, the behavior of
individual elemental ratios can be \it substantially \rm different!

\begin{figure}[htbp]
\centerline{
\psfig{file=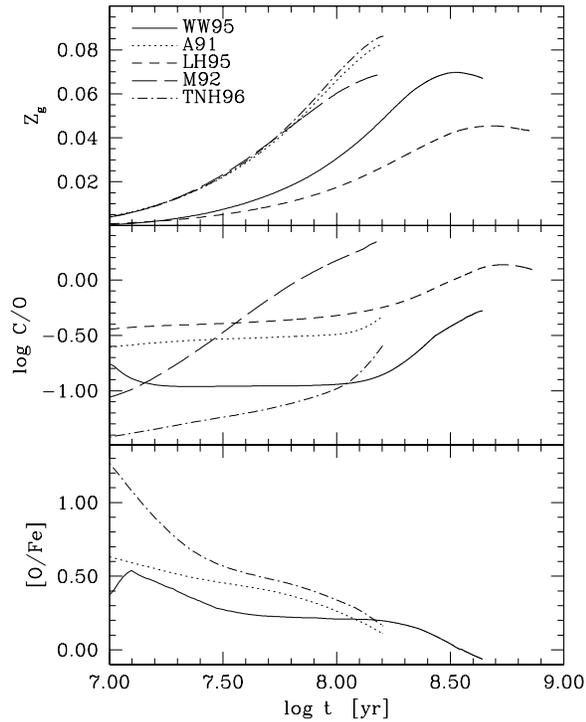,width=3.90in,height=3.90in,angle=0}}
\caption{Evolution of the ISM metallicity Z$_{\rm g}$, C/O, and O/Fe ratios as a
function of time.  The timescale for star formation was varied in order to
recover a (V-K) and luminosity-weighted mean metallicity in agreement with
that observed locally.  Star formation is assumed to cease with the onset of
a galactic wind.}
\end{figure}

\section{Implications}

My only goal in this contribution has been to present a cursory comparison of a
number of Type II SNe yield contributions, demonstrating through simple
non-quantitative figures that there still exist significant differences in the
predictions available in the literature.  One yield package which is entirely
suited to one particular chemical evolution application may be entirely
incompatible with another!  Thomas \etal (1997) have recently come to the same
conclusion.

Finally, we wish to note here 
that the aforementioned uncertainties in the massive star stellar yields
make determining an (accurate)
upper mass limit to the IMF (Gibson 1998) and an (accurate) accounting of the
Type II SNe ICM iron fractionary contribution (Gibson \etal 1997)
a difficult proposition, despite recent claims to the
contrary.

\section{References}
\noindent
Arnett, D. 1991, Frontiers of Stellar Evolution, ed. D.L. Lambert, ASP 

Conf.  Series, 389 (A91)

\noindent
Arnett, D. 1996, Supernovae and Nucleosynthesis, Princeton: Princeton 

Univ.  Press (A96)


\noindent
Gibson, B.K. 1998, ApJ, submitted

\noindent
Gibson, B.K., Loewenstein, M. \& Mushotzky, R.F. 1997, MNRAS, 290, 623

\noindent
Langer, N. 1997, The History of the Milky Way and its Satellite System, 

ed. A.  Burkert, \etal, ASP Conf. Series, 169

\noindent
Langer, N. \& Henkel, C. 1995, Space Sci. Rev., 74, 343 (LH95)

\noindent
Maeder, A. 1992, A\&A, 264, 105 (M92)

\noindent
Reimers, D., \etal 1992, Nature, 360, 561

\noindent
Renzini, A. \& Voli, M. 1981, A\&A, 94, 175 (RV81)

\noindent
Rich, R.M. 1996, New Light on Galaxy Evolution, ed. R. Bender \& R.L. 

Davies, Dordrecht: Kluwer, 19

\noindent
Thielemann, F.-K., Nomoto, K. \& Hashimoto, M. 1996, ApJ, 460, 108 

(TNH96)

\noindent
Thomas, D., Greggio, L. \& Bender, R. 1997, MNRAS, in press

\noindent
Woosley, S.E. \& Weaver, T.A. 1995, ApJS, 101, 181 (WW95)

\end{document}